\documentclass[%
 reprint,
 amsmath,amssymb,
 aps,
]{revtex4-2}

\usepackage{amsmath,amssymb,amsbsy,amstext, amsthm, simplewick, amsfonts}

\usepackage{graphicx}

\usepackage{siunitx}
\usepackage{upgreek}
\usepackage{framed}
\usepackage{wrapfig}
\usepackage{multirow}
\usepackage{bbm}
\usepackage[dvipsnames,table]{xcolor}
\usepackage[colorlinks=true,pdfstartview=FitV,breaklinks=true]{hyperref}
\usepackage{url}
\hypersetup{urlcolor=BlueViolet,
	    citecolor=Plum,
	    linkcolor=PineGreen}

\usepackage[absolute,overlay]{textpos}

\newcommand{\equaref}[1]{Eq.~(\ref{#1})}

\newcommand{\figref}[1]{Fig.~\ref{#1}}

\newcommand{\refref}[1]{Ref.~\cite{#1}}

\begin{document}

\begin{textblock*}{5cm}(18.46cm,1cm)  
    
\end{textblock*}

\preprint{APS/123-QED}


\title{When Tiny Halos Stir Spacetime: Gravitational Waves from Fifth-Force Mergers}

\author{Xinpeng Wang$^{1,2,3}$}
\email{xinpeng.wang{}@ipmu.jp}

\author{Yifan Lu$^{4}$}
\email{yifanlu@ucla.edu}

\author{Zachary S. C. Picker$^{4,5,6}$}
\email{zachary.picker@gmail.com}

\author{Alexander Kusenko$^{1,4}$}
\email{kusenko@ucla.edu}

\author{Misao Sasaki$^{1,7,8,9}$}
\email{misao.sasaki@ipmu.jp}


\affiliation{
 $^1$ Kavli Institute for the Physics and Mathematics of the Universe (WPI), The University of Tokyo Institutes for Advanced Study, The University of Tokyo, Chiba 277-8583, Japan\\
 $^2$ Department of Physics, Graduate School of Science, The University of Tokyo, Tokyo 113-0033, Japan\\
 $^3$ School of Physics Science and Engineering, Tongji University, Shanghai 200092, China\\
 $^4$ Department of Physics and Astronomy, University of California Los Angeles, Los Angeles, California, 90095-1547, USA\\
 $^5$ Department of Physics, Engineering and Astronomy, Queen's University, Kingston ON K7L 3N6, Canada\\
 $^6$ Arthur B. McDonald Canadian Astroparticle Physics Research Institute, Kingston ON K7L 3N6, Canada\\
 $^7$ Center for Gravitational Physics and Quantum Information,
 Yukawa Institute for Theoretical Physics, Kyoto University, Kyoto 606-8502, Japan\\
 $^8$ Leung Center for Cosmology and Particle Astrophysics, National Taiwan University, Taipei 10617, Taiwan\\
 $^9$ Asia Pacific Center for Theoretical Physics, Pohang 37673, Korea 
}




\date{\today}

\begin{abstract}

\noindent Dark matter fermions interacting via attractive fifth forces mediated by a light mediator can form dark matter halos in the very early universe. We show that bound systems composed of these halos are capable of generating gravitational wave (GW) signals detectable today, even when the individual halos are very light. The Yukawa force dominates the dynamics of these halo binaries, rather than gravity. As a result, large GW signals can be produced at initially extremely high frequencies, which are then redshifted to frequency bands accessible to current or future GW observatories. In addition, the resulting GW signals carry distinctive features that enable future observations to distinguish them from conventional ones. Notably, even if only a tiny fraction of dark matter experiences strong fifth-force interactions, such effects provide a new avenue to discover self-interacting dark matter through GW observations.
\end{abstract}

\maketitle

\noindent The nature of dark matter (DM) remains one of the most compelling mysteries in modern physics. While the cold DM paradigm successfully explains the large-scale structure of the universe, the microscopic properties and possible non-gravitational interactions of DM remain elusive. Self-interacting dark matter (SIDM) has emerged as a well-motivated candidate, capable of addressing small-scale structure anomalies \cite{Moore:1994yx,Moore:1997sg, Klypin:1999uc, Moore:1999nt, Vogelsberger:2012ku} and potentially leaving observable imprints in cosmic history~\cite{Amendola:1999er,Farrar:2003uw, Spergel:1999mh,Pollack:2014rja}. 

In particular, an intriguing possibility arises when the DM is (partly) composed of dark fermions that experience additional long-range attractive forces through light mediators, such as scalar bosons involved in Yukawa coupling. This mediator could even be associated with a dynamical dark energy (DE) component, for which the DESI collaboration has recently reported strong evidence~\cite{DESI:2024mwx, DESI:2025zgx}. Such attractive forces, when much stronger than gravity, can render the DM fermion number density perturbations unstable in the early universe, leading to the rapid collapse of small overdensities into structures and virialized dark halos~\cite{Savastano:2019zpr, Domenech:2021uyx,Domenech:2023afs} even during radiation domination---a process otherwise almost impossible under purely gravitational dynamics. Subsequent radiative cooling through the mediator in asymmetric DM models~\cite{Petraki:2013wwa, Petraki:2013wwa, Zurek:2013wia,Barr:1990ca} can further drive these halos to form compact objects like fermi balls or (primordial) black holes~\cite{Amendola:2017xhl,Flores:2020drq,Flores:2021jas, Flores:2023zpf, Lu:2024xnb}.

Regardless of their final fate, the formation of intermediate fermionic DM halos is inevitable. Previously, the production of gravitational waves (GWs) from the aspherical collapse of these individual halos was considered~\cite{Flores:2022uzt}. In this work, we investigate the GW signals from halo binaries. If two halos form in close proximity, they can become bound to each other by the Yukawa force. If a third halo is present to add a tidal force to the two halos, they can form a binary, producing GWs as they inspiral. Remarkably, those GW signals, although generated during the very early universe, could be strong enough to be detected by existing and future GW observatories, opening a new window into the physics of DM and interactions between DM and DE.

In this \textit{letter}, we investigate the GW signals from DM halo binaries formed during radiation domination via a long-range attractive force much stronger than gravity, using the Yukawa interaction as a concrete example.  By combining with the thermal evolution of the halos, we explore the possible evolutionary pathways of such binaries and identify the resulting GW signatures that could be observed today. 

\textit{Binary halos with Yukawa forces.}
The power of GW is proportional to the square of the third time derivative of the mass quadrupole moment, $P_{\mathrm{GW}}\propto \dddot{Q}_{ij} \dddot{Q}^{ij}$, or equivalently to the sixth power of the angular frequency of the orbit $P_{\mathrm{GW}}\propto w^6$. 
Thus, any mechanism capable of significantly accelerating the orbital motion can strongly enhance the GW luminosity. In the standard scenario of compact binaries composed of neutron stars or black holes, the orbital dynamics are almost entirely governed by gravity. This is because compact objects are effectively electrically neutral, leaving gravity as the dominant force.
However, if an additional long-range attractive force exists---for instance, arising from new physics beyond the Standard Model---the orbital acceleration can be substantially amplified. Intuitively, a force stronger than gravity would drive the two bodies to orbit at higher velocities, increasing the higher-order time derivatives of the quadrupole moment and boosting the GW power spectrum beyond the predictions for binaries bound solely by gravity.
Motivated by this possibility, we investigate fermionic dark halos interacting through additional long-range forces as promising candidates to realize such scenarios. 

As a concrete example, we consider an attractive force mediated by a scalar field $\phi$, which couples to a fermion $\psi$ through a Yukawa interaction:
\begin{equation}
    \mathcal{L}\supset\frac{1}{2}m_\phi^2\phi^2+m_\psi\bar{\psi}\psi-y\phi\bar{\psi}\psi+\cdots,
\end{equation}
where we assume there is some asymmetry in the heavy fermion $\psi$ and that the mediator is either massless or very light, $m_{\phi}\ll m_{\psi}$, with the reduced Planck mass $M_\mathrm{pl}=(8\pi G)^{-1/2}=2.43\times 10^{18}\mathrm{GeV}$. 
Throughout this work, we assume that the fermions $\psi$ remain non-relativistic during the epoch of interest. This condition can be easily achieved by restricting to the epoch where the temperature of the universe satisfies $T \ll m_{\psi}$. 
This framework is particularly well-motivated since it can naturally trigger dark halo formation during radiation domination~\cite{Savastano:2019zpr,Flores:2022uzt}, provided that the condition $z/z_\mathrm{eq}\ll 12\beta^2$ is satisfied before the force is effectively screened at $z_{\mathrm{end}}\approx\left(\Omega_{\mathrm{rad},0}H_0^2l_{\mathrm{y}}^2\right)^{-1/4}$ (See Appendix~\ref{haloform} for a detailed discussion). Here, $l_{\mathrm{y}}$ is the effective range of the Yukawa force, and $\beta$ parameterizes its relative strength compared with gravity,
\begin{equation}
    \beta^2\equiv\frac{y^2M_\mathrm{pl}^2}{m^2_\psi},
\end{equation}
which is taken to be constant in the following. Given the parameters above, we can estimate the maximal halo mass, given by the total mass of $\psi$'s enclosed within $l_{\mathrm{y}}$ when it re-enters the horizon, i.e. $H^{-1}\sim l_{\mathrm{y}}$,
\begin{equation}
    M_{\mathrm{y}}= 2\pi f_{\psi}\left(\frac{a_{\mathrm{end}}}{a_{\mathrm{eq}}}\right)M_{\rm pl}^2l_{\mathrm{y}},
    \label{my}
\end{equation}
where $f_\psi=\rho_\psi/(\rho_m+\rho_\psi)$ is the fraction of $\psi$ in in the total matter component.
In the radiation dominated era, the $\psi$ halos can rapidly form and pair up due to the strong force, leading to a population of binaries.
The Yukawa interaction results in an attractive force between the charges in the virialized clumps of $\psi$ (hereinafter referred to as `halos').
 Two halos carrying net scalar charges $q_1$ and $q_2$ with masses $m_1$ and $m_2$ separated by a distance $d$ experience an attractive potential:
\begin{equation}
    V=-\frac{1}{d}\left(Gm_1m_2+\frac{y^2q_1q_2}{4\pi}e^{-m_\phi d}\right)~.
    \label{potential}
\end{equation}
If the separation is much smaller than the effective range $d\ll l_{\mathrm{y}}=m_{\phi}^{-1}$, the exponential factor $e^{-m_\phi d}$ becomes negligible, and the potential becomes Coulomb-like:
\begin{align}
    V=-\frac{1}{d}Gm_1m_2(1+2\beta^2),
    \label{effectivepotential}
\end{align}
implying the orbital motion of the halos approximately follows Keplerian dynamics. 
For two halos with a separation $d$ to decouple from the Hubble flow and form a bound system, the following condition must be satisfied:
\begin{equation}
    \mu H^2d^2-\frac{Gm_1m_2}{d}-\frac{y^2 q_1q_2}{d}<0.
\end{equation}
where $\mu$ is the reduced mass and $H$ is the Hubble parameter.
Therefore, when $2\beta^2\gg 1$, 
two halos which would otherwise remain unbound under pure gravity can form bound system due to the Yukawa forces when ${Gm_1m_2}(1+2\beta^2)>\mu H^2d^3$ is achieved. Consequently, the presence of the additional attractive Yukawa force can significantly enhance the binary formation rate during radiation domination.
\begin{table*}
    \centering
    \begin{tabular}{c|c|c}
    \hline
    &Gravitationally-Bound & Yukawa-Bound \\
    \hline
   Strain $h_z$& 
      $ \frac{4}{d_{L}c^4}\left({\pi f_z }\right)^{2/3}(G\mathcal{M}_z)^{5/3}\quad(\star)$&$
        (\star)\times \left(2\beta^2\right)^{2/3}$
  \\
  \quad&\quad&\quad
  \\
    Frequency $f_z=f/(1+z)$ & $\frac{1}{\pi(1+z)}\sqrt{\frac{G(m_1+m_2)}{a^3}}\quad(\circ)$&$
    (\circ)\times\sqrt{2\beta^2}$
  \\
  \quad&\quad&\quad
  \\
    Orbital Evolution ${df_z}/{dt_z}$& $\frac{96}{5}\pi^{8/3}\left(\frac{G\mathcal{M}_z}{c^3}\right)^{5/3}f_z^{11/3}\quad(\triangle)$&$(\triangle)\times\left(2\beta^2\right)^{5/3}$
  \\
    \hline
    \end{tabular}%
    \caption{The properties for the redshifted GW emitted at $z$ from circular orbits in the inspiral phase in the limit of $2\beta^2\gg 1$, where $\mathcal{M}_z=(1+z)\mathcal{M}=(1+z)(m_1m_2)^{3/5}/(m_1+m_2)^{1/5}$ is the redshifted chirp mass, $f_z\equiv 2f_0/(1+z)$ is the redshifted GW frequency, and the $d_{L}\equiv(1+z)\int_{0}^z \mathrm{d}z' H(z')^{-1}$ is the luminosity distance.}
    \label{gwstrain}
\end{table*}
Once a binary halo system forms within the Yukawa range, its dynamics are well described by Keplerian motion with a modified orbital frequency $2\pi f_{0}=\sqrt{{G(m_1+m_2)(1+2\beta^2)}{a^{-3}}}$, where $a$ is the semi-major axis of the orbit. However, for binary halos with a separation comparable to or greater than the effective range $l_{\mathrm{y}}$, the screening of the force strongly suppresses the additional binding energy arising from the scalar force, making stable binary formation unlikely in this regime. In those cases, for the highly eccentric orbits, if the pericenter lies in the effective range, the Yukawa force can induce a rapid increase in eccentricity. Moreover, since the Yukawa force does not follow an exact $1/d^2$ scaling, it causes significant apsidal precession already at the 0th-order of post-Newtonian expansion (0PN), with precession angle approximate by $\Delta\varpi\approx 2\pi(\sqrt{[1+a/l_{\mathrm{y}}]/[1+a/l_{\mathrm{y}}-(a/l_{\mathrm{y}})^2]}-1)$ for nearly-circular orbits with $2\beta^2\gg 1$ \footnote{Absence of the Yukawa force, the apsidal precession during one period of revolution is given $\Delta\theta\approx \frac{6\pi GM}{a(1-e^2)c^2}$ (1PN)}. Consequently, the appearance of the Yukawa force in a two-body system will leave distinctive low-frequency features on the waveform of GW in comparison to the gravity-only case, which is a smoking gun signature of the theory.
Examples and explanation of these effects, obtained by numerically solving a two-body system, are shown in Appendix~\ref{twobody}. In the following analysis, we restrict our attention to the simplest case where the binary separation remains entirely within the effective range of the force and the orbit is Keplerian (either the eccentricity $e=0$ or $a/l_{\mathrm{y}}\ll 1$ 
), allowing the effective potential to be accurately described by \equaref{effectivepotential}. 
In analogy to the electromagnetic radiation from a Coulomb two-body system, both scalar and gravitational waves are emitted from the binary system due to the non-vanishing third time derivatives of its scalar and mass quadrupole moments. Notably, unlike in the electromagnetic case, dipole radiation is always absent here because the two bodies carry the same kind of scalar charges,
rendering the total scalar dipole moment constant in time. We can compute the power of the scalar radiation as well as the gravitational radiation using the quadrupole formula,
\begin{equation}
\begin{aligned}
&P_{\mathrm{SW}}=\left(\frac{y^2}{4\pi}\right)^4\frac{32}{5 a^5}\frac{(m_1+m_2)(m_1m_2)^2}{m_\psi^8}\sum_{n=1}^{\infty}\mathcal{I}_Q(e,n),\\
&P_{\mathrm{GW}}=\left(\frac{y^2}{4\pi}\right)^3\frac{32Gm_\psi^2}{5a^5}\frac{(m_1+m_2)(m_1m_2)^2}{m_\psi^8}\sum_{n=1}^{\infty}\mathcal{I}_Q(e,n).\\
\end{aligned} 
\end{equation}

The radiation power spectrum of a stable periodic orbit consists of a set of $\delta$-functions peaked at the discrete harmonics, $f_n=nf_0 \ ~(n\geq 2)$, and the subscript $z$ throughout refers to the redshifted frequency $f_z=f/{(1+z)}$ as would be detected at present. For near-circular orbits, the power spectrum reaches its maximum at $n_{\mathrm{peak}}=2$, whereas for highly eccentric orbits, the power is concentrated at higher harmonics with $n_{\mathrm{peak}}>2$ (see Appendix~\ref{radiation} for a detailed derivation). In the strong force limit, the redshifted GW signatures for a circular orbit are summarized in Table \ref{gwstrain}. We find that the strain, frequency, and orbital evolution of the GWs are enhanced by different powers of $\beta$ without altering the scaling with the observed frequency. As a result, there is an interesting degeneracy between the GW signals of these events and with those from conventional low-redshift, massive sources. Similar ideas have been explored in the context of dark binaries with additional long-range interactions \cite{Bai:2023lyf,Bai:2024pki,Kopp:2018jom} without considering quadrupole scalar radiation. 

To compare with the sensitivity curves of the current GW detectors, we write down the characteristic strain for different harmonics detected today, assuming that they are emitted from an individual halo binary with initial eccentricity $e$ at redshift $z$,
\begin{equation}
\begin{aligned}
 h_{c,z}^{n}&=\frac{1}{\pi d_{L}}\sqrt{{2G} \frac{dE_{\mathrm{GW},n}}{d{f_{z,n}}}}\\
&=(2\beta^2+1)^{5/6}\sqrt{\frac{2}{3}}\frac{\left({G\mathcal{M}_z}\right)^{5/6}}{d_{L}\pi^{2/3}}(f_{z,n})^{-1/6}\\
&\quad\times \left(\frac{\mathcal{I}_{Q}(e,n)}{(n/2)^{2/3}F(e)}\right)^{1/2}.
\end{aligned}
\end{equation}
For a concrete example, we can examine a binary in a circular orbit at $z=10^{9}$, with rest frame chirp mass $\mathcal{M}=10^{-5}M_{\odot}$, comoving frame frequency $f_{z,2}=10^{-3}\mathrm{Hz}$, fermion mass $m_{\psi}=10^{11}\mathrm{GeV}$, and Yukawa coupling constant $y=0.1$. Then we obtain a characteristic strain $h_c\approx 5\times 10^{-18}$, which falls in the sensitivity curve of LISA ~\cite{LISA:2017pwj}. Thus, even \textit{individual} halo binaries in the very early universe, not just the stochastic signal, can be detectable today.

Let us now estimate the stochastic GW background from these halos. We will assume the inspirals are identical and isotropically distributed within a relatively narrow redshift bin $\delta z \ll z$ shortly after the halo formation epoch $z \sim z_{\mathrm{end}}$, so that the mean separation of the population is $\bar{d} \sim l_{\mathrm{y}}$. Before the Hubble horizon approaches the scale $l_{\mathrm{y}}$, the fermion density is too high to allow the formation of a large number of isolated binary halos. Instead, structures and halos emerge primarily through collective, N-body–like interactions, during which the particle motions remain largely coherent. As a result, in the absence of significant non-spherical initial conditions, no substantial GW emission from halo binaries is expected during this structure formation phase. Consequently, the dominant contribution to the stochastic gravitational wave background (SGWB) arises from halo binaries after formation. Binary systems can form efficiently when $\bar{d}$ becomes larger than $l_{\mathrm{y}}$, allowing individual binaries to evolve and merge without being significantly perturbed by nearby halos. 
Since there is no significant formation of new halos due to the collapse after $l_{\mathrm{y}}$ enters the horizon, most newly formed halos arise from mergers of pre-existing ones. 
As these mergers proceed, the overall halo number density decreases and $\bar{d}$ increases, suppressing further binary formation. 
When $\bar{d}$ becomes much larger than $l_{\mathrm{y}}$, halo interactions become negligible and the system effectively freezes out into isolated mergers or remnant halos.

The characteristic strain of the SGWB is thus given by~\cite{Phinney:2001di}:
\begin{equation}
\begin{aligned}
h_{c,z}^{\mathrm{sto}}&=\sqrt{\frac{4G}{\pi f_{z}^2}\int_{0}^{\infty} N(z)f_{z}\frac{dE_{\mathrm{GW}}}{df}\mathrm{d}z}\\
    &\approx(2\beta^2+1)^{5/6}\sqrt{\frac{N(z)\delta z}{(1+z)^{1/3}}}\frac{2\left({G\mathcal{M}_z}\right)^{5/6}}{\sqrt{3}\pi^{1/6}}f_z^{-2/3},
\end{aligned}
\label{hsto}
\end{equation}
where $N(z)\delta z$ is the number of merger events per unit comoving volume within the redshift bin of width $\delta z$. 
We give a rough estimation to $N(z)\delta z$ in Appendix~\ref{numberevent}, which shows that as $\bar{d}/l_{\mathrm{y}}$ increases due to binary coalescence or Hubble expansion, the binary formation rate~\equaref{prob} is strongly suppressed by force screening, validating the narrow redshift bin argument.
While the above provides a rough estimate of the stochastic background, it is worth noting that this does not preclude the possibility of rare, high-SNR individual merger events. Similar to the LIGO case, where individual binary coalescences are observed but the stochastic background remains undetected, rare low-redshift or high-mass halo mergers might clearly stand out above the integrated background.

\begin{figure}[!htbp]
\centering
\includegraphics[width=0.5\textwidth]{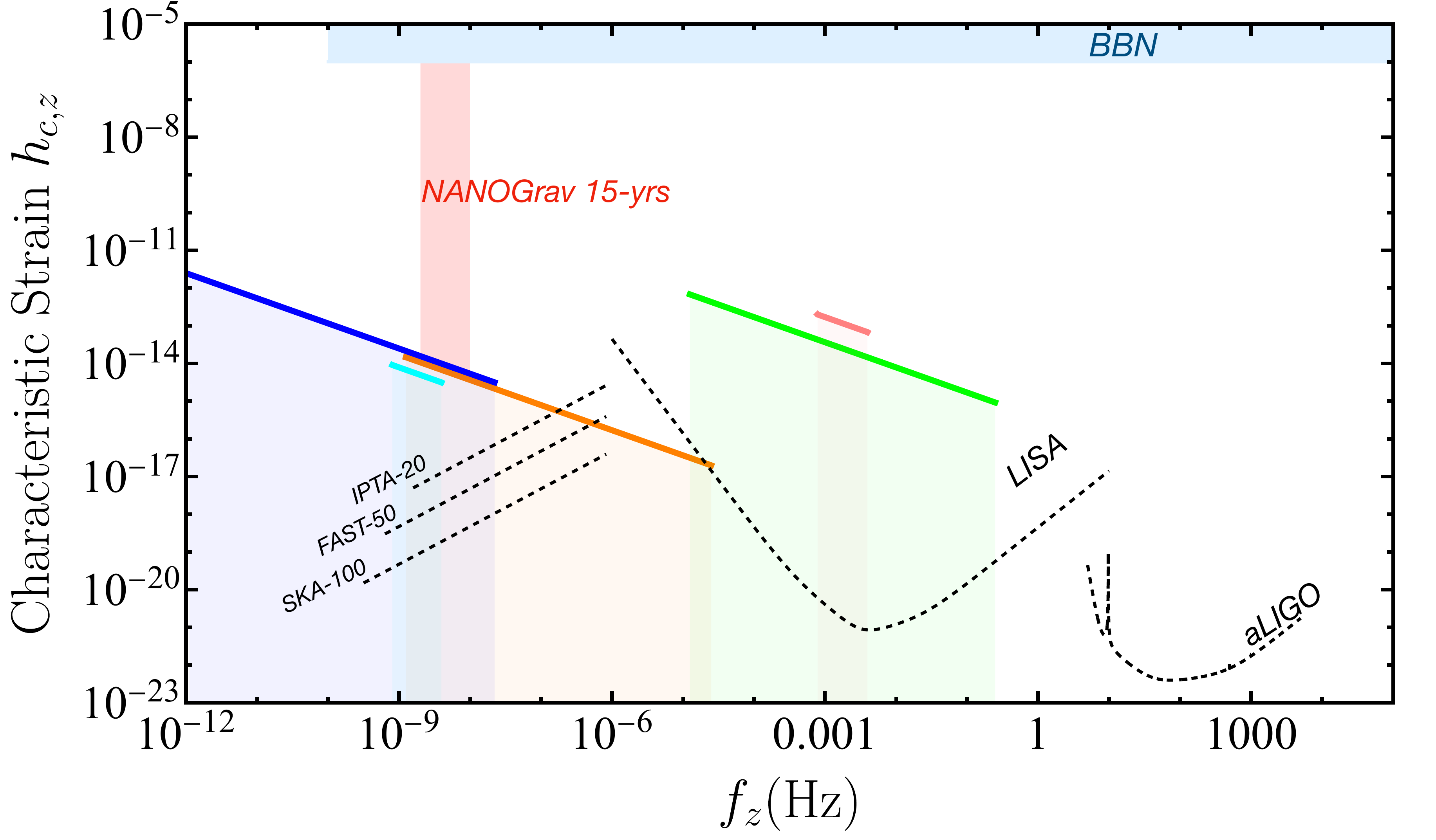}
\caption{The characteristic strain of the SGWB from halo mergers is plotted, alongside sensitivity curves of both current and planned GW surveys and constraints. Parameters are fixed at $y=0.1$ and $m_{\mathrm{h}}=10^{-10}M_{\mathrm{y}}$ (as in Fig.~\ref{parameterspace1}), with $(m_\psi, m_\phi)$ corresponding to the colored dots in Fig.~\ref{parameterspace1}. In the plot, we consider a total comoving merger density $R(z)\delta z=3\times 10^{-6}(M_{\odot}/m_{\mathrm{h}})/\mathrm{Gpc}^3$.
Sensitivity curves are taken from PTAs \cite{Kuroda:2015owv}, LISA \cite{Robson:2018ifk}, and aLIGO \cite{LIGO_T1800044_v5_2018}. The light-red region shows the signal and constraints from the NANOGrav best-fit SIGW from BSMBH 
inspirals \cite{NANOGrav:2023gor}. The light-blue region is the constraint on the SGWB coming from Big Bang Nucleosynthesis (BBN).}
\label{hcz}
\end{figure}

\begin{figure}[!htbp]
\centering
\includegraphics[width=0.5\textwidth]{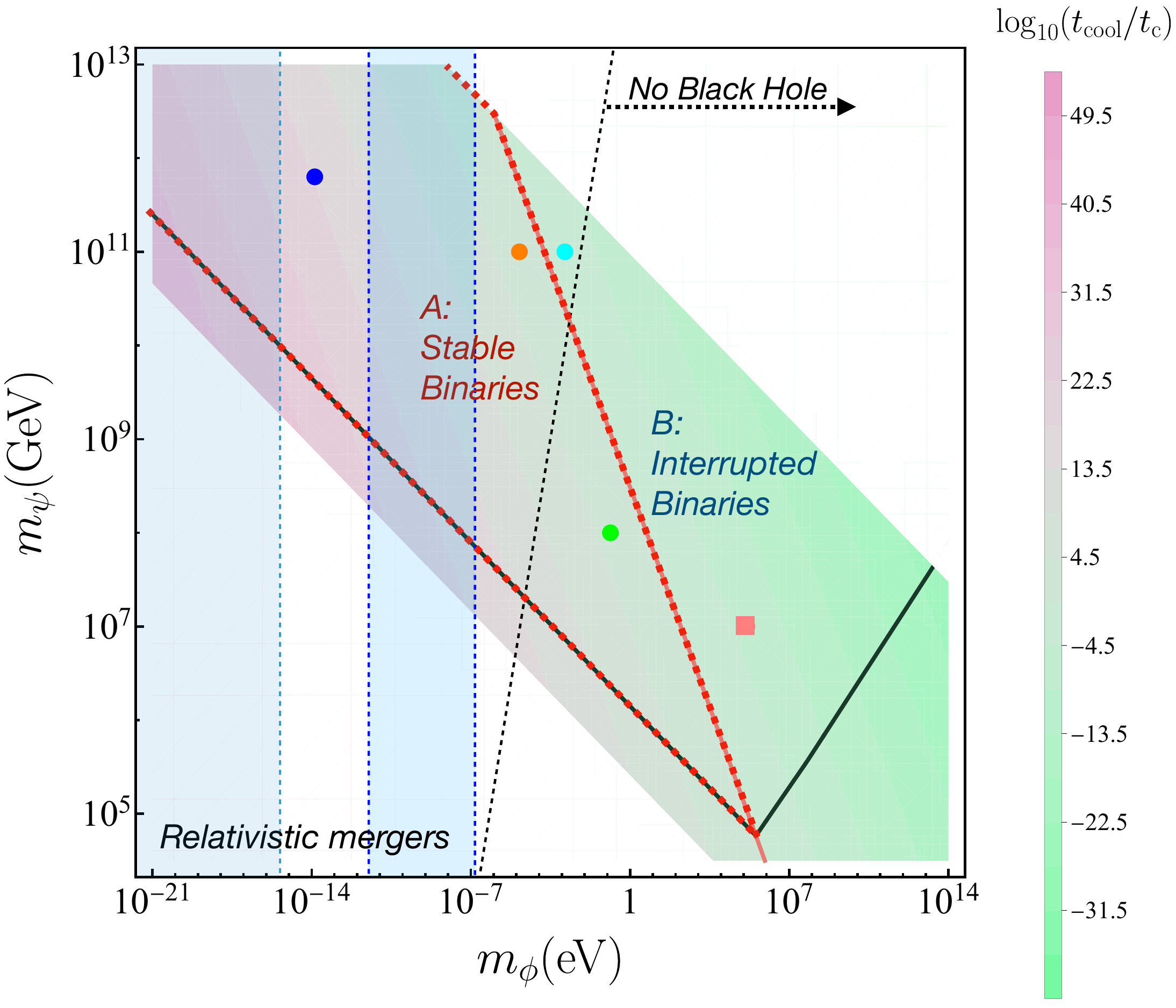}
\caption{
Contour of the time ratio $t_{\mathrm{cool}}/t_\mathrm{c}$ for a Yukawa-bound binary halo. 
We fix $y=0.1$ and $m_1=m_2=m_h=10^{-10}M_{\mathrm{y}}$, where $M_{\mathrm{y}}$ is the total mass of $\psi$ fermions enclosed within the Yukawa range $l_{\mathrm{y}}$ at the end of halo formation, defined in Eq.~(\ref{my}). 
The initial orbit is circular with separation $d=l_{\mathrm{y}}$ and halo radius $R=l_{\mathrm{y}}(m_{\mathrm{h}}/M_{\mathrm{y}})^{1/3}$. 
The red solid line denotes $t_{\mathrm{cool}}=t_{\mathrm{c}}$, with $t_{\mathrm{cool}}>t_\mathrm{c}$ ($<t_\mathrm{c}$) on the left (right). 
Stable inspirals satisfying $\min(t_{\mathrm{cool}},t_\mathrm{c})>T_0$ occupy the region above the black curve; 
category~A systems ($t_{\mathrm{cool}}>t_\mathrm{c}>T_0$) are enclosed by red dashed lines. 
White areas are excluded where assumptions break down, and the blue-shaded region is ruled out by observational constraints from Bullet Cluster (left), and $\gamma$-ray and BBN on evaporating PBHs (right). 
The black dashed contours indicate where halos cannot collapse into black holes without additional mass growth, since the Schwarzschild radius would be smaller than the Compton wavelength of $\psi$.
The square marker denotes a case with uncertain post-cooling effective charge retention: the final state could either retain most of its effective charge or lose it significantly, depending on the underlying theory. 
In \figref{hcz}, we treat this case as in Category~B.
A more detailed analysis will be presented in future work.
}
\label{parameterspace1}
\end{figure}

For the SGWB produced in this case, taking $z=10^{15}$ and $m_\psi=10^{11}\mathrm{GeV}$, the corresponding halos with chirp masses $\mathcal{M}=1.2\times 10^6\mathrm{g}$ result in a characteristic strain $h_{c,z}^{\mathrm{sto}}\approx 2\times10^{-15}$ at $f_{z}\equiv f_{z,2}=1\mathrm{yr}^{-1}$, which is able to explain or at least contribute significantly to the NANOGrav 15-year observation of the SGWB. In \figref{hcz}, we present the characteristic strain corresponding to the parameter choice described above, along with additional parameter sets shown as colored dots in \figref{parameterspace1}. The signal frequencies span a wide range, from as low as $10^{-12}\mathrm{Hz}$ up to around $1\mathrm{Hz}$, making them potentially detectable by pulsar timing array surveys such as IPTA~\cite{Hobbs:2009yy,10.1063/1.2900317,EPTA:2023fyk,Jenet:2009hk,NANOGrav:2023gor,Manchester:2012za,Reardon:2023gzh, Nobleson:2021ngl}, CPTA~\cite{Lee2016_ASPC502_19,Xu:2023wog} and SKA~\cite{Lazio:2013mea,Spiewak:2022btk}, as well as space-based interferometers like LISA~\cite{LISA:2017pwj}, Taiji~\cite{Hu:2017mde} and TianQin~\cite{TianQin:2015yph}.


\textit{From halos to black holes via radiative cooling.}
As discussed in Ref~\cite{Flores:2023zpf}, the halos composed of Yukawa fermions can collapse into stable compact objects, such as degeneracy pressure-supported Fermi balls, nuggets, or black holes, after a phase of efficient radiative cooling. For halos with a large amount of charges, their fate ultimately depends on the parameters of the theory, which requires a Tolman–Oppenheimer–Volkoff (TOV)/Chandrasekar-type analysis. The critical mass in this scenario has been studied in detail in \refref{Gresham:2017zqi, Gresham:2018rqo, Xie:2024mxr, Lu:2024xnb}. According to the no-hair theorem \cite{Carter:1971zc, Ruffini:1971bza, Hawking:1971vc, Robinson:1975bv}, a stationary black hole cannot retain its scalar charges upon collapse. 
Notably, even for a dense Fermi ball, the effective scalar charge can be strongly suppressed, as the scalar field acquires a larger effective mass inside the ball, leading to enhanced screening of the interaction  \cite{Lu:2024xnb}. 
In a bound two-body system, such an abrupt loss of charge eliminates the associated scalar-mediated attractive force between the two bodies. If this force had contributed significantly to the system's binding energy, its sudden disappearance could cause the system to become unbound and eject the bodies at high speed. Using energy conservation, the ejection speed can be approximated by
$v_{\mathrm{rel}}\sim \sqrt{4G(m_1+m_2)\beta^2(d_{\mathrm{e}})^{-1}}$, where $d_{\mathrm{e}}$ is the separation when the ejection happens. A population of ultra-fast compact objects, e.g., free-floating planets, could also behave as anomalously short microlensing events, offering a potential observational signature of this scenario~\cite{2011,Niikura:2017zjd,Mr_z_2018,Niikura:2019kqi,Mr_z_2019,DeRocco:2023hgh,DeRocco:2023gde}.

Hereinafter, we revisit the thermal history of the halo after virialization~\cite{Flores:2020drq}.  At first, when the fermion number density inside of the halo remains relatively low, the fermions can be treated as individual incoherent sources of scalar radiation and the radiated power can be estimated from $P_{\mathrm{incoh}}\sim (y^2)^3q^3/(4m^2R^4)$ for a halo having $q$ particles within the radius $R$. 
As the fermion system becomes denser, i.e. when $R\lesssim y^2q/m_\psi$ (such that the interaction energy becomes comparable to or larger than the fermion kinetic energy) the scalar radiation can be estimated analogously to photon bremsstrahlung in a plasma, with power $P_\mathrm{brem}\sim (y^2)^4q^8/(m_\psi^2R^4)\ln(y^2q/(m_\psi R))$. 

The radius of the halo will shrink until the scalar diffusion time scale becomes longer than the fermions' free-fall time scale, ${R} \lesssim y^2q^{3/5}/m_\psi$, at which point the scalar radiation becomes trapped inside the halo. At this stage, the halo becomes optically thick and quickly turns into a fireball of temperature $T_\mathrm{halo}\sim \sqrt{yq}/R$. The cooling then proceeds by the black body radiation with power $P_\mathrm{surf}\sim4\pi (y^2)q^2/R$ until the halo contracts into a degenerate Fermi ball or black hole. Therefore, we can naively estimate the halo cooling time scale by
\begin{equation}
\begin{aligned}
    t_\mathrm{cool}\approx\frac{E_\mathrm{h}}{P_{\mathrm{incoh}}+P_{\mathrm{brem}}+P_{\mathrm{surf}}},
\end{aligned}   
\end{equation}
where $E_\mathrm{h}\sim y^2q^2/R$ is total energy of
the halo.

Typically, the cooling dominated by incoherent radiation and bremsstrahlung is much longer than the surface cooling time. Meanwhile, the strong scalar wave emission will cause the binary to lose energy more rapidly, significantly reducing the coalescence time. It is then possible for the coalescing time to be comparable to, or longer than, the halo cooling time.  For the orbit with an initial eccentricity $e$ near to 0, the coalescence time is given by
\begin{equation}
    t_\mathrm{c}\approx\frac{5}{256}\frac{c^5 a^4}{G^3m_1m_2(m_1+m_2)}(2\beta^2+1)^{-3},
\end{equation}
which can be derived by accounting for the energy loss due to both scalar radiation and gravitational radiation (see detailed analysis in Appendix \ref{coalescingtime}).

By comparing the cooling time $t_\mathrm{cool}$, the coalescing time $t_\mathrm{c}$, and the orbital period $T_0$, the fate of the Yukawa-bound halo binary formed in the early universe can be roughly classified into one of three potential outcomes.
\begin{itemize}
    \item A. \textit{Stable Binary.} If $t_\mathrm{cool}\gg t_\mathrm{c}\gg T_0$, the binary can survive for many orbital periods without merging. Continuous GWs and scalar waves (SWs) are stably emitted over a long inspiral phase. A final burst of GWs and SWs occurs at the end of the chirp when the merger takes place. The inspiral signal terminates at the innermost stable circular orbit (ISCO), which naively is related to the radii of the halos. This is an important feature in the GW spectrum, which would allow us to distinguish the halo merger scenario from Binary Supermassive Black Hole (SMBH) mergers. The parameter space corresponding to this outcome is surrounded by red dashed curve in Fig.~\ref{parameterspace1}.
    \item B. \textit{Interrupted Binary.} If $t_\mathrm{c}\gg T_0$ is satisfied but $t_\mathrm{cool}\ll t_\mathrm{c}$, the binary is initially stable but is disrupted when the Yukawa-mediated attraction disappears abruptly (e.g., due to scalar charge loss as a result of halo collapse into a black hole or fermi ball.). Continuous GW and SW emissions are seen during the early stage, but the sudden force decrease can eject the newly formed compact objects at high velocities, which would not happen in normal primordial black hole (PBH) scenarios.
    In this case, the inspiral terminates earlier than expected. The cut-off frequency is related to the nearest separation of the halos. As shown in Fig. \ref{hcz} and \ref{parameterspace1}, for examples that lie in the parameter space of category B (purple and pink), the band widths of the SGWB spectra are much narrower compared to the complete inspirals.
   
    \item C. \textit{Prompt Merger.} If $t_\mathrm{c}<T_0$, the binary halos merge directly before completing a full revolution. This usually implies that the halo velocities are relativistic. During this process, instead of a long chirp with many cycles, the GW and SW have a burst-like profile. 
    
\end{itemize}

\textit{}

\textit{Constraints and Discussions.}
Constraints on the parameters $m_{\psi}$ and $m_{\phi}$ are assembled with the time ratio contour on the $(m_\psi,m_\phi)$ plane in Fig.~\ref{parameterspace1}. In the right-up corner, the parameter space is excluded since the criterion of the halo formation \equaref{criterion} cannot be satisfied.
Notably, the contour in the left-down corner is absent because the analytical results for the time ratios no longer apply for relativistic mergers. However, in this corner, short-time GW bursts can be emitted in prompt merger cases. This phase, which is usually highly relativistic, should be solved through Numerical Relativity (NR). Here we estimate the total energy of GW during the burst by considering a simplified process where a halo is thrust towards another, by $\Delta E\sim\beta^4\gamma^2(G^3m_1^2m_2^2/R^3)$ in the ultrarelativistic limit, where  $\gamma=1/\sqrt{1-v^2}$ is the Lorentz factor and $R$ is the radius of the halo (See detailed analysis for the prompt merger case in Appendix~\ref{radiation}).
Similar to a gamma ray burst, the GW energy is predominantly emitted along the motion of the particle with an opening angle $|\theta|<\gamma^{-1}$.

Observationally, strong DM self-interactions are tightly constrained by systems such as the Bullet Cluster and large-scale structure observations when the self-interacting component constitutes the dominant DM fraction. The most relevant bound arises from the Bullet Cluster on the momentum-transfer cross section~\cite{Markevitch:2003at}:
\begin{align}
    \frac{\sigma}{m}\,f_{\mathrm{DM}} \lesssim 1~\mathrm{cm^2\,g^{-1}} \simeq 4.7\times10^3~\mathrm{GeV^{-3}},
\end{align}
where we take $f_{\mathrm{DM}} \approx f_{\psi}$ as the fractional abundance of $\psi$ in DM. 
Assuming that all $\psi$ particles are bound within Fermi balls with an effective cross section $\sigma\sim\pi m_{\phi}^{-2}$, this limit sets a lower bound on $m_{\phi}$, as shown in \figref{parameterspace1}.
Our results further indicate that even strongly interacting dark sectors coupled via a fifth force can be probed through gravitational-wave signals from halo mergers, 
even when $\psi$ constitutes only a small fraction $f_{\psi}\ll1$ of the total DM, since $f_{\psi}$ only mildly shifts the viable parameter space by modifying the maximum halo mass Eq.~(\ref{my}). 
In addition, if all fermi ball DM eventually collapses into black holes, we apply $\gamma$-ray and BBN constraints on evaporating black holes with masses $10^{9}\sim 10^{16}\,\mathrm{g}$, as shown in \figref{parameterspace1}.



\textit{Conclusion.} We have identified a novel mechanism for generating detectable gravitational waves (GWs) in the early universe from dark halo binaries bound by attractive long-range fifth forces. The presence of such forces significantly enhances binary formation during radiation domination and boosts the GW luminosity beyond the purely gravitational scenario. We predict both distinct stochastic and individual-event signatures, potentially lying in the detectable bands of ongoing and planned PTAs surveys and LISA, offering an unexplored observational window to probe macroscopic dark matter (DM). These results provide a concrete target for upcoming GW observations and open a promising avenue---even a tentative detection of such non-standard signatures in GWs would provide a unique probe of DM at cosmological scales. 










\begin{acknowledgments}
\noindent We thank Ying-li Zhang for the discussions during the early stage of the work. 
We would also like to acknowledge Robert Brandenberger, Elisa G. M. Ferreira, Anamaria Hell, Jinsu Kim, Kazuya Koyama,  Xiao-Han Ma, Tom Melia, Masahiro Takada, and Tsutomu T. Yanagida for their useful comments on the work.

This work is supported in part by JSPS KAKENHI Grant No. JP24K00624 [M.S.], by Forefront Physics and Mathematics Program to Drive Transformation (FoPM), a World-leading Innovative Graduate Study (WINGS) Program, the University of Tokyo [X.W.], and by the U.S. Department of Energy (DOE) Grant No. DESC0009937 [Y.L., Z.P., and A.K.]. Kavli IPMU is supported by the World Premier International Research Center Initiative (WPI), MEXT, Japan.
X.W. acknowledges the insightful inputs from the 3rd International Workshop on Gravitational Waves and the Early Universe (GWEU25) co-organized by Ningbo University, ITP-CAS, and the Center for High Energy Physics, Peking University.

\end{acknowledgments}
\bibliographystyle{bibi}

\bibliography{apssamp}

\onecolumngrid
\section*{Supplemental Material}
\appendix
\section{The Scalar and Gravitational Radiation}
\label{radiation}
\subsection{Non-Relativistic System}
In the following, we compute the non-relativistic scalar radiation mainly following \cite{LandauLifshitz_TheClassicalTheoryOfFields}. 
The result is useful for both the bound and unbound system.
Working in the center-of-mass frame, considering two point charges $q_1$ and $q_2$ with masses $m_1$ and $m_2$, each of the charges is located at $(d_1\cos\varphi,d_1\sin\varphi)$ and ($d_2\cos\varphi,d_2\sin\varphi$) with a separation of $d=d_1+d_2$, where
\begin{equation}
\begin{aligned}
d_1=\frac{m_2}{m_1+m_2}d,\quad d_2=\frac{m_1}{m_1+m_2}d.
\end{aligned}
\end{equation}
As we know, for two point charges attracting each other following the Coulomb law, they form elliptical orbit. The equation of motion for the orbit is given by 
\begin{equation}
        \frac{p}{d}=1+e\cos\xi
\end{equation}
where $p=a(1-e^2)$ is the semi positive focal length, $a$ the semi major axis and $e$ is the eccentricity.
We decompose the equation of motion to  parametric equations for the coordinates $\mathbf{x}=\mathbf{d}\cos\phi$ and $\mathbf{y}=\mathbf{d}\sin\phi$
\begin{equation}
\begin{aligned}
    x&=a(\cos\xi-e),\\
    y&=a\sqrt{1-e^2}\sin\xi,\\
    \omega_0t&=\xi-e\sin \xi,
\end{aligned}    
\end{equation}
where the fundamental angular frequency and the period of the orbit is given
\begin{equation}
\begin{aligned}
     \omega_{0}=&\sqrt{\frac{y^2 q_1 q_2}{4\pi\mu a^3}}\\
     T_0=&2\pi\sqrt{\frac{4\pi\mu a^3}{y^2 q_1q_2}},
\end{aligned}  
\end{equation}
if assuming $2\beta^2\gg1$.
As a result, we obtained the Fourier components of the coordinates
\begin{equation}
\begin{aligned}
     &x_n=\frac{i}{\omega_0nT}\int_0^Te^{i\omega_0nt}\dot{x} \mathrm{d}t=\frac{a}{n}J_n'(ne),\\
    &y_n=\frac{i}{\omega_0nT}\int_0^Te^{i\omega_0nt}\dot{y} \mathrm{d}t=\frac{ia\sqrt{1-e^2}}{2\pi n e}J_{n}(ne).
\end{aligned}
\end{equation}
The mass dipole and quadrupole moments as well as the charge dipole and quadrupole moments of the system is therefore given by
\begin{equation}
\begin{aligned}
        &\mathbf{D}_{q}=\sum_{i} q_i\mathbf{d}_i,\\
        &\mathbf{D}_{m}=\sum_{i} m_i\mathbf{d}_i,\\
        &{Q}_{q}^{ab}=\sum_{i}q_i\left(3d_i^ad_i^b-d^2\delta^{ab}\right),\\
        &{Q}_{q}^{ab}=\sum_{i}m_i\left(3r_i^ar_i^b-d^2\delta^{ab}\right).
\end{aligned}
\end{equation}
For the point two point charges evolving in an arbitrary elliptical orbit with eccentricity $e$, the scalar dipole radiation intensity $I\equiv 2\pi dE/(d\omega)$ peaked at the discrete harmonics, $\omega_{n}=n\omega_0$. Therefore, the intensity for in the Fourier space for the  $n^{\mathrm{th}}$ harmonic reads
\begin{equation}
\begin{aligned}
      I_{D,n}&=\frac{y^2}{4\pi}\frac{4\omega_0^4 n^4}{3c^3}| \mathbf{D}_{q,n}|^2,\\&
       =\left(\frac{y^2}{4\pi}\right)^3\frac{ (q_1q_2)^2 n^2}{  3a^4}\left(\frac{q_1}{m_1}-\frac{q_2}{m_2}\right)^2\mathcal{I}_{D}(e,n),
\end{aligned}
\end{equation}
where
\begin{align}
    \mathcal{I}_{D}(e,n)\equiv 4\left(J_{n}^{'2}(ne)+\frac{1-e^2}{e}J_{n}^{2}(ne)\right)
\end{align}
is normalized to be 1 when $n=1$ for circular orbits. If we consider $m_i/q_i=m_\psi$ which is a constant, the dipole radiation vanishes just like the gravitational radiation. Under this assumption, the center-of-mass frame is equivalent to the center-of-charge frame.

Using the same method, the scalar quadrupole radiation intensity for $n^{th}$ harmonic is given by
\begin{equation}
\begin{aligned}
        I_{Q,n}&=\frac{y^2}{4\pi}\frac{\omega_0^6 n^6}{90\pi }|Q_{q,n}^{ab}|^2\\
        &=\left(\frac{y^2}{4\pi}\right)^4\frac{8(q_1q_2)^3\mu n^2}{5 a^5}\left(\frac{q_1}{m_1^2}+\frac{q_2}{m_2^2}\right)^2\mathcal{I}_{Q}(e,n)\\
\end{aligned}
\end{equation}
where 
\begin{equation}
\begin{aligned}
\mathcal{I}_{Q}(e,n)=&\left.\frac{n^4}{32}\right\{[J_{n-2}(ne)-2eJ_{n-1}(ne)\\
&\frac{2}{n}J_n(ne)+2eJ_{n+1}(ne)-J_{n+2}(ne)]^2\\
&+(1-e^2)\left[J_{n-2}(ne)-2J_n(ne)+J_{n+2}(ne)\right]^2\\
&\left.+\frac{1}{3n^2}[J_n(ne)]^2\right\},
\label{quadrupole}
\end{aligned}
\end{equation}
and 
\begin{align}
    F(e)=\sum_{n=1}^{\infty}\mathcal{I}_{Q}(e,n)=\frac{1+(73/24)e^2+(37/96)e^4}{(1-e^2)^{7/2}}.
\end{align}

Due to the fact that the power spectrum of radiation is a set Dirac $\delta$ functions peaked at the harmonics, the total power is given by the simply summing up from $n=1$ to $\infty$,
\begin{align}
P_{\mathrm{SW}}=\sum_{n=1}^{\infty}I_{D,n}+I_{Q,n}+O(I_{S,n})
\end{align}

For circular orbit $e=0$, $\mathcal{I}_{Q}(e,n)$ is only non-vanishing when $n=2$. For the elliptical orbit case, the radiation intensity peaks in higher harmonics. Considering that $m_i/q_i=m_{\psi}$, the scalar radiation is dominated by lowest order ($2^{nd}$) harmonic quadrupole radiation. The power of the scalar radiation is
\begin{equation}
    \begin{aligned}
P_{\mathrm{SW},n}&\approx\left(\frac{y^2}{4\pi}\right)^4\frac{32}{5 a^5}\frac{(m_1+m_2)(m_1m_2)^2}{m_\psi^8}\mathcal{I}_{Q}(e,n).
    \end{aligned}
\end{equation}
Similarly, we can also estimate the energy loss caused by GW radiation,
\begin{equation}
    P_{\mathrm{GW}}=\frac{G}{90c ^5}|\dddot Q_{g}^{ab}|^2.
\end{equation}
The Fourier component of the radiation intensity,
\begin{equation}
P_{\mathrm{GW},n}=\frac{G\omega_0^6n^6}{90}|Q_{g}^{ab}|^2.
\end{equation}
Replacing the quadrupole mass moment $Q_{g,n}^{ab}$ by
\begin{equation}
\begin{aligned}
     |Q_{g,n}^{ab}|&=\frac{(m_1+m_2)m_1m_2}{m_2^2q_1+m_1^2 q_2}    |Q_{q,n}^{ab}|\\&=m_\psi|Q_{q,n}^{ab}| \quad(\mathrm {if}\ q_i=m_i/m_\psi).
\end{aligned}
\end{equation}
As a result, the power of the quadrupole GW is
\begin{align}
        P_{\mathrm{GW},n}=\left(\frac{y^2}{4\pi}\right)^3\frac{32G}{5a^5}\frac{(m_1+m_2)(m_1m_2)^2}{m_\psi^6}\mathcal{I}_{Q}(e,n).
\end{align}
We also obtain the ratio of the gravity case to the scalar case
\begin{align}
    \frac{P_{\rm GW,2}}{P_{\rm SW,2}}\propto\left(\frac{4\pi Gm_\psi^2}{y^2}\right)=\frac{1}{2\beta^2}.
\end{align}

\subsection{Relativistic System}
For the prompt merger cases explained in the main text, we estimate the total energy of GW emission by simplifying the scenario to: gravitational radiation emitted by a charged particle with mass and charge ($m_2$, $q_2$) thrusted into a halo with size $R$, mass and charge ($m_1$, $q_1$). The particle has a finite kinetic energy at the initial time. We assume flat space time since the size of the halo is much larger than the Schwarzschild radius $(R\gg 2GM)$. 
According to \cite{Ruffini:1981af}, for a particle moving along $z$-axis with 4 velocity given $u^{\mu}=\gamma(1,0,0,v)$, the retarded gravitational potential perturbation is given in linearized Einstein equations
\begin{equation}
    \psi^{\mu\nu}=-4m_2\left[\frac{u^{\mu} u^\nu}{|\mathbf{x}-\mathbf{z}(s)|l_{\alpha}u^{\alpha}}\right]_{\mathrm{ret}}
\end{equation}
where $\mathbf{x}$ is the field point and $\mathbf{z}(s)$ is the position of the particle at its retarded time $z^0(s)$, $l_{\alpha}=(1,(\mathbf{x}-\mathbf{z}(s))/|\mathbf{x}-\mathbf{z}(s)|)$ is the pointing vector from the particle to the field point.
The non-vanishing terms are
\begin{equation}
\begin{aligned}
     &\psi_{00}=4m_2\left[\frac{\gamma}{|\mathbf{x}-\mathbf{z}(s)|(1-v\cos\theta)}\right]_{\mathrm{ret}},\\
     &\psi_{0z}=4m_2\left[\frac{\gamma v}{|\mathbf{x}-\mathbf{z}(s)|(1-v\cos\theta)}\right]_{\mathrm{ret}},\\
     &\psi_{zz}=4m_2\left[\frac{\gamma v^2}{|\mathbf{x}-\mathbf{z}(s)|(1-v\cos\theta)}\right]_{\mathrm{ret}},
\end{aligned}
\end{equation}
where the $\cos\theta$ is the $z$-component of $l_\alpha$.
For a field point that is far away from the two body system, we introduce, $r=|\mathbf{x}|\approx|\mathbf{x}-\mathbf{z}(s)|$ as the distance between the field point. The TT component of $\psi_{\mu\nu}$, which is the strain of the GW is thus given:
\begin{equation}
\begin{aligned}
    &\psi^{TT}_{\theta\phi}=0,\\
    &\psi^{TT}_{\theta\theta}=\psi^{TT}_{\phi\phi}=\frac{2m_2}{r}\left[\frac{\gamma v^2\sin^2\theta}{1-v\cos\theta}\right]_{\mathrm{ret}},
\end{aligned}
\end{equation}
which are the components of the orthonormal diad basis $(e^{\theta},e^{\phi})$.
The effective energy-momentum tensor of the gravitational wave is therefore given
\begin{equation}
    (T_{\mathrm\mu\nu})_{\mathrm{GW}}=\frac{1}{32\pi G}\langle\psi^{TT}_{\alpha\beta,\mu}(\psi^{TT})^{\alpha\beta}_{,\nu}\rangle=\frac{Gm_2^2}{4\pi r^2}\langle A^2l_{\mu}l_{\nu}\rangle,
\end{equation}
where 
\begin{equation}
    A=\sin^2\theta\gamma v \dot v[1+\gamma^2(1-v\cos\theta)]/(1-v\cos\theta)^3.
\end{equation}

\begin{figure*}[!htbp]
\centering
\includegraphics[width=0.5\textwidth]{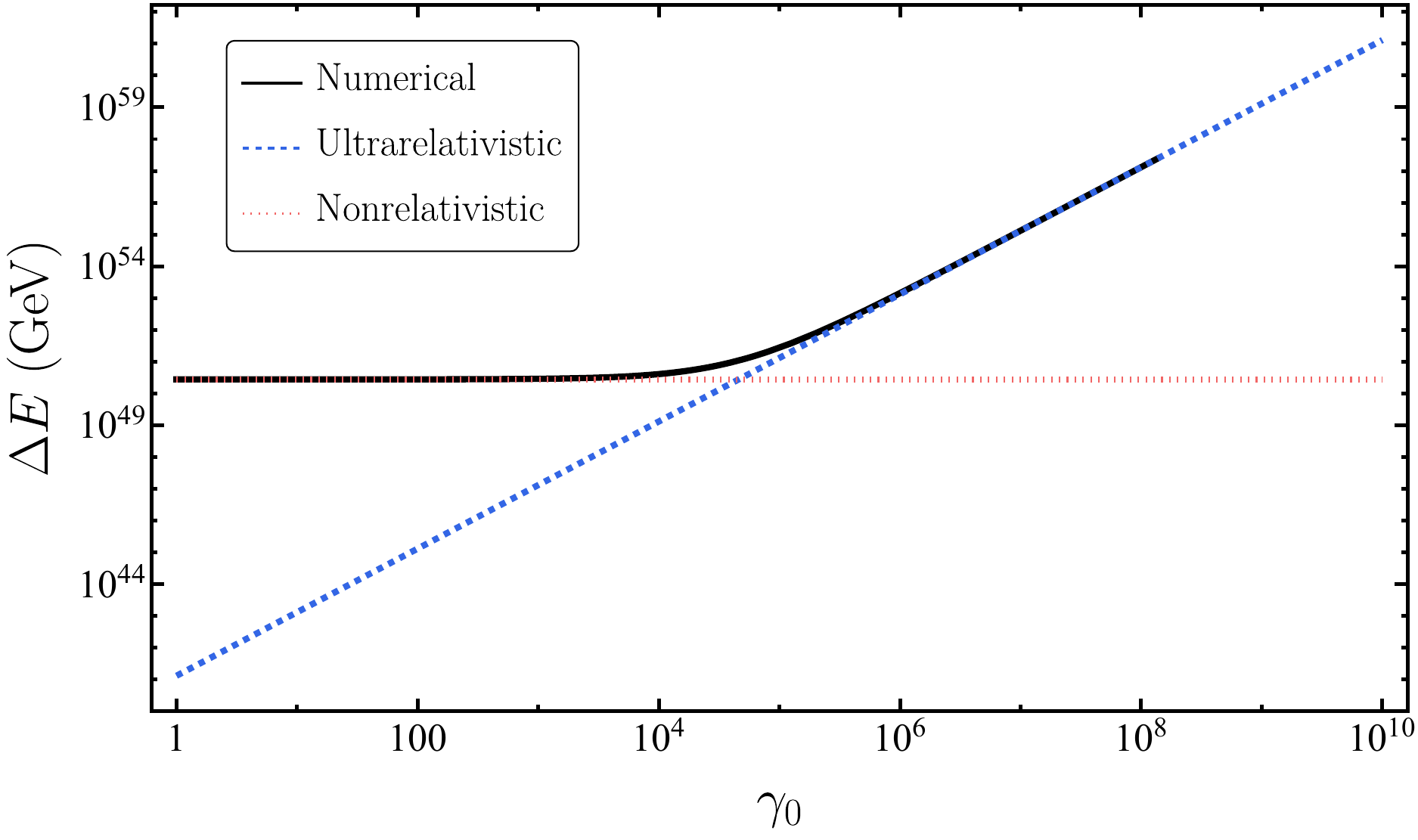}
\caption{Total gravitational--wave energy $\Delta E$ from the prompt merger calculated using \equaref{deltaEresult}. 
The plot is obtained for halo masses $m_1=m_2=2\times10^{17}\,\mathrm{g}$, mediator mass $m_\psi=m_1/q_1=10^{10}\,\mathrm{GeV}$, halo radius $R=3\times10^{-3}\,\mathrm{m}=2\times10^{-10}Gm_1$, initial separation $d_0=\infty$, and coupling constant $y=0.1$. 
The black solid curve shows the result if full numerical integral, the blue dashed curve the ultrarelativistic approximation, and the red dotted curve the non-relativistic approximation.}
\label{deltaEgamma}
\end{figure*}

Thus, the total energy emitted during the process is obtained by integrating the GW energy flux $l^aT_{0a}$ through a sphere of radius $r$ and over the emission time interval
\begin{equation}
    \Delta E=\int_{0}^{t_{\mathrm{end}}}\mathrm{d}t\int \mathrm{d}\Omega \ r^2 l^{a}(T_{0a})_\mathrm{GW}=\frac{G m_2^2}{2}\int_{0}^{\infty}\mathrm{d}t\int_{0}^{\pi} \mathrm{d}\theta \sin\theta A^2(\theta,t).
    \label{deltaE}
\end{equation}
Consider the acceleration due to the Yukawa force in the $m_1$'s rest frame, the $i$ component of the four acceleration is given
\begin{align}
   \frac{d|u^{i}|}{d\tau}=\gamma^4 \frac{d v}{d t}=\gamma\frac{y^2q_1q_2}{4\pi m_2 d(t)^2}
\end{align}
where $\tau$ is the proper time, $dt/d\tau=\gamma$.
Therefore, the acceleration of $m_2$ deep inside of the Yukawa potential is given by
\begin{align}
    \frac{d v}{d t}=-\frac{y^2q_1q_2}{4\pi \gamma^3 m_2 d(t)^2}.
    \label{dvdt}
\end{align}
In the following, we do the integral \equaref{deltaE} step by step. We first do the integral over $\theta$ to obtain the power of the GW emission,
\begin{align}
    P=\frac{Gm_2^2}{2}\int_{0}^{\pi} \mathrm{d}\theta \sin\theta A^2(\theta,t)=\frac{4\pi G}{15}\left(\frac{y^2 q_1q_2}{4\pi}\right)^2\gamma^2\frac{4v^2+5+15v^{-2}+15\tanh^{-1}(v)(v-v^{-3})}{d(t)^4}.
\end{align}
The next step is to solve for the evolution of $v$ as a function of $d$ in order to do the integral over time interval. According to \equaref{dvdt}, we obtain
\begin{equation}
    \gamma(t)-\gamma_0=\frac{y^2 q_1q_2}{4\pi}\frac{1}{m_2}\left(\frac{1}{d(t)}-\frac{1}{d_0}\right),
\end{equation}
where $\gamma_0$ and $d_0$ are the Lorentz factor and separation at the initial time.
As a result, we obtain the total energy emitted during the collision
\begin{equation}
    \Delta E=\int_{d_{0}}^{R}P(d)v^{-1}(d)\mathrm{d}d=\frac{4\pi G}{15}\left(\frac{y^2 q_1q_2}{4\pi}\right)^2I(d_0,\gamma_0)
    \label{deltaEresult}
\end{equation}
where we show the asymptotic results in ultrarelativistic and nonrelativistic limit respectively,
\begin{equation}
    I(d_0,\gamma_0)=\left\{
    \begin{aligned}
        &8\gamma_0^2\left(\frac{1}{R^3}-\frac{1}{d_0^3}\right)\quad \mathrm{when\ }\gamma_0\gg 1,\\
        &\frac{3}{10m_2^2}\left(\frac{y^2q_1q_2}{4\pi}\right)^2\left(\frac{1}{R^5}-\frac{1}{d_0^5}\right)\quad\mathrm{when}\ \gamma_0\rightarrow1, 
    \end{aligned}
    \right.
\end{equation}
By doing the numerical integral for an arbitrary initial condition, we show the relation between $\Delta E$ and $\gamma_0$ in Fig. \ref{deltaEgamma}.

\begin{figure*}[!htbp]
\centering
\includegraphics[width=0.5\textwidth]{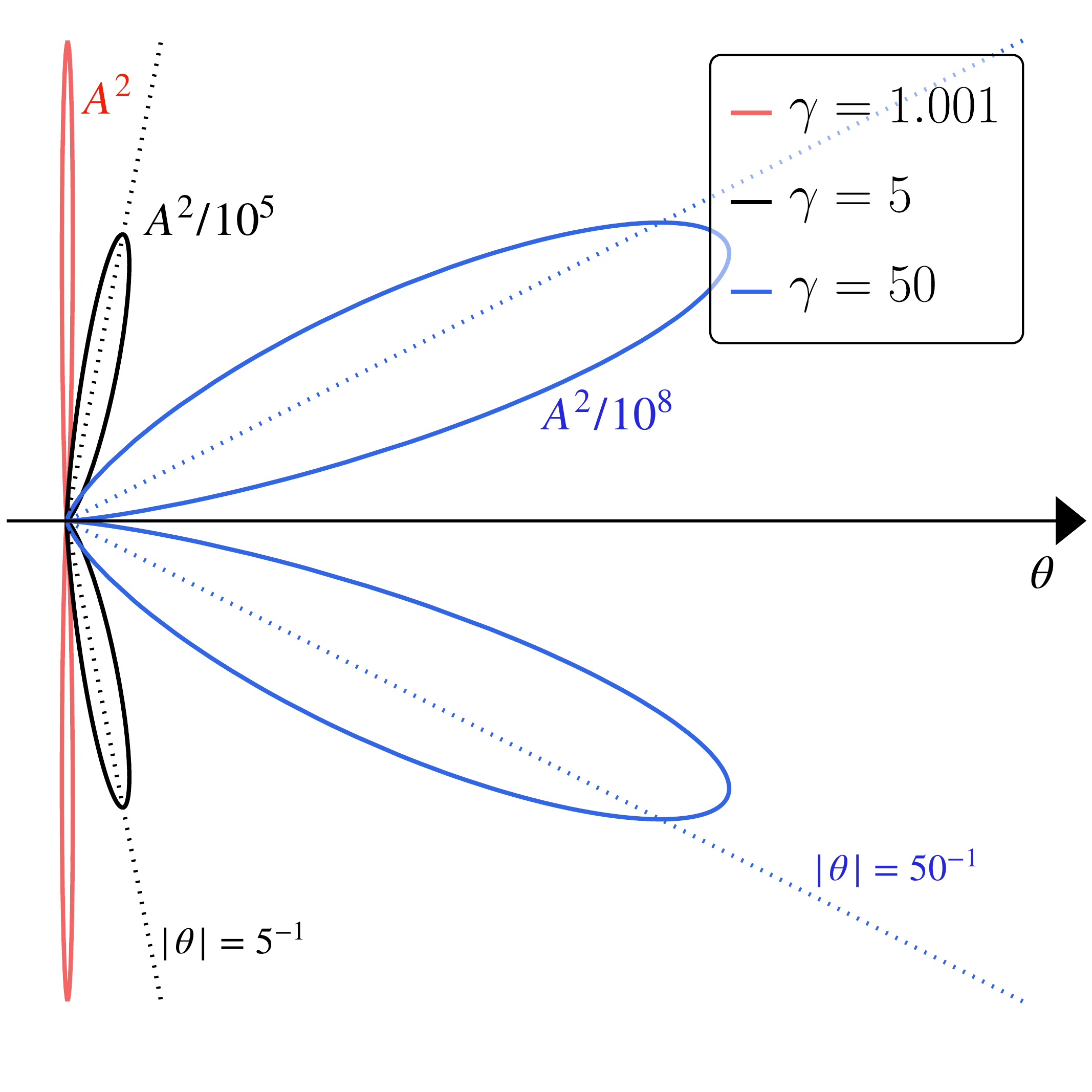}
\caption{Relativistic beaming effect in gravitational--wave burst scenarios. The contours show the angular distribution of the emitted power, $\mathrm{d}P/\mathrm{d}\Omega\propto A^2$, for different Lorentz factors $\gamma$.}
\label{angle}
\end{figure*}

Another feature that potentially interesting for the GW signal from the prompt merger case is that in the relativistic limit, the energy is predominantly emitted along the motion of the particle with an opening angle $\Delta\theta<\gamma^{-1}$, just like the relativistic beaming effect in gamma ray burst, as shown in Fig. \ref{angle}.

\section{The Coalescing Time and Orbital Evolution}
\label{coalescingtime}
In the limit of $2\beta^2\gg 1$, the energy loss of the system is dominated by scalar radiation. We compute the coalescing time of the binary dominated by Yukawa force according to \cite{PetersMathews_1963}. For simplicity, we focus on the circular orbit and neglect the background Hubble flow. 
The total energy of the orbit is given by
\begin{equation}
    E=-\frac{y^2}{4\pi}\frac{q_1q_2}{2a}.
    \label{totalenergy}
\end{equation}
Assuming Keplerian orbit with an initial eccentricity $e$, the radiation power due to the scalar quadrupole radiation is given
\begin{equation}
\begin{aligned}
    \frac{dE}{dt}&=-\frac{y^2}{4\pi}\frac{q_1q_2}{2a^2}\frac{da}{dt}\\
&=-\left(\frac{y^2}{4\pi}\right)^4\frac{32(q_1q_2)^3\mu }{5  a^5}\left(\frac{q_1}{m_1^2}+\frac{q_2}{m_2^2}\right)^2F(e)
\label{energyloss}
\end{aligned}
\end{equation}
By solving the above equation given the boundary conditions $a_\mathrm{i}=a$ and $a_\mathrm{f}=0$, we obtain the coalescing time in the $e\rightarrow0$ limit
\begin{equation}
  t_\mathrm{c}=\frac{5}{256}\left(\frac{4\pi}{y^2}\right)^3\frac{a^4}{(q_1q_2)^2\mu}\left(\frac{q_1}{m_1^2}+\frac{q_2}{m_2^2}\right)^{-2}.
\end{equation}
The ratio between the coalescing time and the fundamental period of the binary system is thus given
\begin{equation}
    \frac{t_\mathrm{c}}{T_0}\sim \frac{5}{512\pi}\left(\frac{4\pi}{y^2}\right)^{5/2}\frac{a^{5/2}}{(q_1q_2)^{3/2}\mu^{3/2}}\left(\frac{q_1}{m_1^2}+\frac{q_2}{m_2^2}\right)^{-2}
\end{equation}

At same time, we can derive the orbital evolution based on the same method. The frequency of the GW from a circular orbit
\begin{align}
    f&=2f_0=\frac{1}{\pi}\sqrt{\frac{y^2 q_1q_2}{4\pi\mu a^3}}\\
    &\rightarrow a=\frac{1}{\pi^{2/3} f^{2/3}}\left(\frac{y^2 q_1q_2}{4\pi \mu}\right)^{1/3}.
\end{align}
We rewrite the \equaref{energyloss} in terms of GW frequency evolution
\begin{equation}
\begin{aligned}
\frac{df}{dt}=\left(\frac{y^2}{4\pi G}\right)^{5/3}\left(\frac{q_1}{m_1^2}+\frac{q_2}{m_2^2}\right)^2\left(\frac{q_1q_2}{m_1m_2}\right)^{2/3}\mu^2\times \frac{df_{G}}{dt},
\end{aligned}
\end{equation}
where in the pure gravity case for a eccentric orbit,
\begin{equation}
     \frac{df_{G}}{dt}=\frac{96}{5}\pi^{8/3}\left(\frac{G\mathcal{M}}{c^3}\right)^{5/3}f^{11/3} F(e),\quad(\triangle)
\end{equation}
where the chirp mass
\begin{equation}
    \mathcal{M}\equiv\frac{(m_1m_2)^{3/5}}{(m_1+m_2)^{1/5}}.
\end{equation}

Under the assumption of $q_i=m_i/m_\psi$, we obtain
\begin{equation}
       \frac{{df}/{dt}}{{df_{G}}/{dt}}=(2\beta^2)^{5/3}. 
\end{equation}
As a result, the energy spectrum of the GW for the $n^{th}$ harmonic is given
\begin{equation}
\begin{aligned}
        &\frac{dE_{\mathrm{GW},n}}{df_{z,n}}=\frac{P_{\mathrm{GW},n}}{(n/2)df_{z}/dt_z}\\&=(2\beta^2)^{5/3}\frac{1}{3}\pi^{2/3}G^{2/3}\mathcal{M}_z^{5/3}f_{z,n}^{-1/3}\frac{\mathcal{I}_{Q}(e,n)}{(n/2)^{2/3}F(e)}.
\end{aligned}
\end{equation}

\section{Two Body Simulation}
\label{twobody}
To account for the screening effect, we perform a non-relativistic 2D two-body simulation of the potential given by \equaref{potential}, neglecting the energy loss due to radiation.
By fixing the $2\beta^2=10^4$ and $m_1=m_2$, we consider 3 types of initial conditions:
\begin{itemize}
 
    \item I. The semi-major axis $a$ and perigee $a(1-e)$ of the orbit are much larger than the effective range of the Yukawa interaction: (a) $a=30m_\phi^{-1}$, $e=0$ and (b)  $a=30m_\phi^{-1}$, $e=0.5$. In this case, the Yukawa force is screened.
    \item II. The semi-major axis $a$ is less than the effective range of the Yukawa interaction: (a) $a=\frac{1}{2}m_\phi^{-1}$, $e=0$, and  (b) $a=\frac{1}{2}m_\phi^{-1}$, $e_\mathrm{i}=0.5$. In this case, the orbit stays in the field of effective range of the Yukawa force.
    \item III. The semi-major axis $a$ is larger than the effective range of the Yukawa interaction while the perigee is smaller than the effective range: (a) $a=2m_\phi^{-1}$, $e=0.5$ and (b) $a=2m_\phi^{-1}$, $e=0.9$. In this case, the tracks will travel in and out of the effective range.    
\end{itemize}

To compute the strain of the GW, we use the formula
\begin{equation}
\begin{aligned}
         &\frac{h_{+}d_{L}}{G/c^2}=\frac{d^2}{dt^2}\left(Q_{xx}-Q_{yy}\right),\\
         &\frac{h_{+}d_{L}}{G/c^2}=2\frac{d^2}{dt^2}\left(Q_{xy}\right),\\
         &Q_{xx}=\mu x^2, Q_{yy}=\mu y^2, Q_{xy}=\mu xy.
\end{aligned}
\end{equation}

\begin{figure*}[!htbp]
\centering
\includegraphics[width=0.8\textwidth]{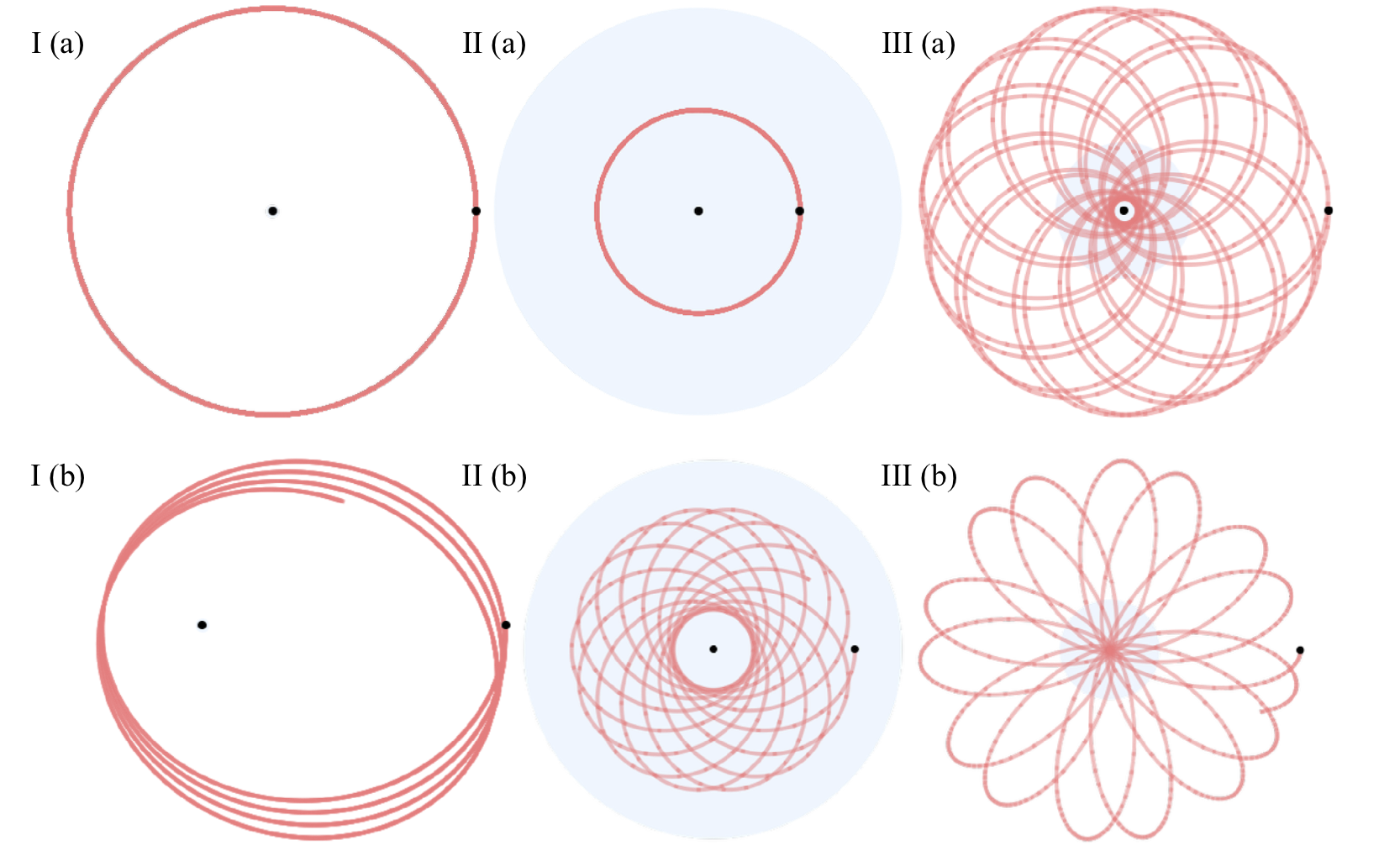}
\caption{The trajectory of the black hole in a bound system for different types of initial conditions. The blue-shaded area shows the effective range of the Yukawa force. The black dots show the initial position of the two halos, with one halo fixed at the origin. The red lines show the trajectory of another halo on the plane.}
\label{trajectory}
\end{figure*}

\begin{figure*}[!htbp]
\centering
\includegraphics[width=1.\textwidth]{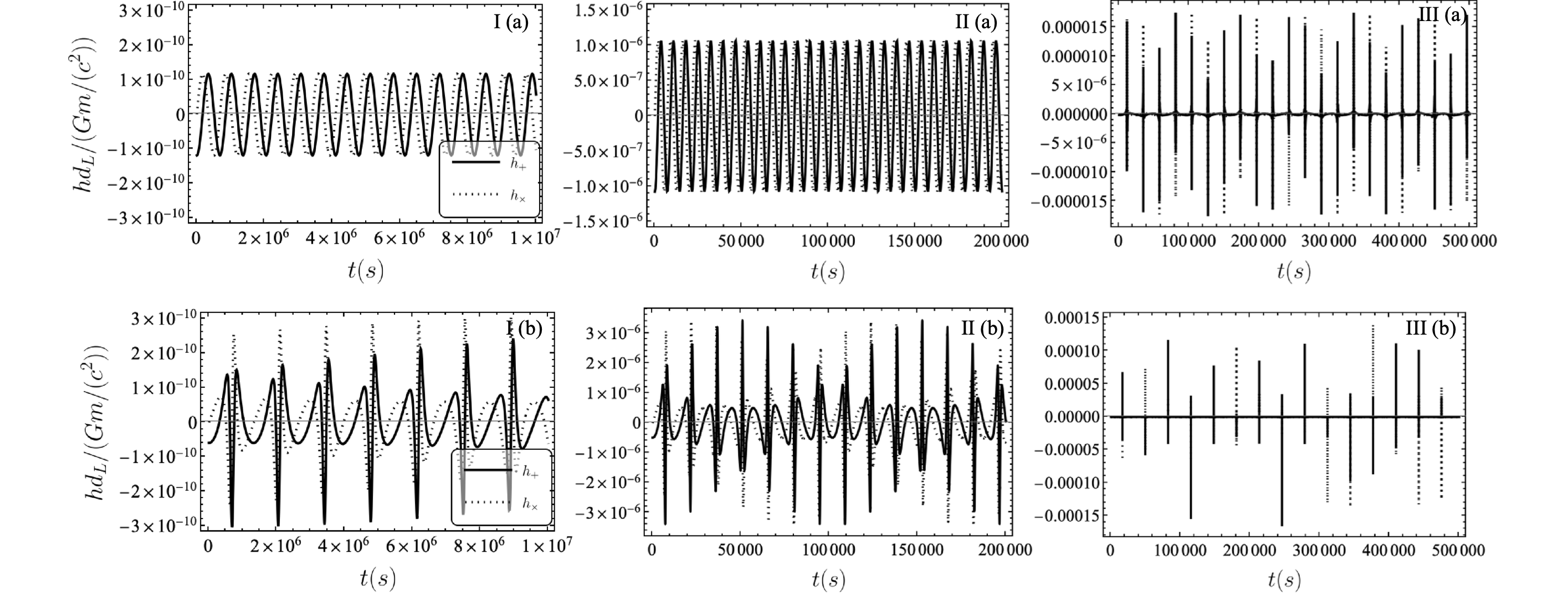}
\caption{The normalized inspiral waveform of the GW with the same setup and time duration as plotted in Fig.~\ref{trajectory}.}
\label{strain}
\end{figure*}

Figure \ref{trajectory} clearly shows that orbits with a non-zero initial eccentricity always exhibit apsidal precession. In Case I, the precession advances only by a very small angle per orbital cycle, since the Yukawa force is effectively screened along the orbit. In contrast, in Cases II and III, the precession proceeds with a much larger angular shift per cycle.


In cases II and III, the quadrupole radiation is greatly enhanced when the halo enters the effective range of the Yukawa force and approaches the halo located at the origin (see Fig.~\ref{strain}). For case III, even though the halo binary starts outside the effective range, the initial eccentricity will be enlarged by the attractive force, causing the halo to pass by the central halo at an extremely fast speed and at a very short distance. As a consequence, the GWs emitted in II and III have their frequency as well as the strain enhanced. The simplest circular orbit case is given in Table~\ref{gwstrain}.

\section{An estimation to the halo binary number density}
\label{numberevent}
Assuming the mass distribution of the halos is monochromatic and follow Poisson distribution in space, we estimate the formation rate of the halo binaries after their formation. Let's focus on the case that the halo pairs could form binary due to a third halo that induces angular momentum also existing within the effective range and coalesce rapidly. The mechanism was studied in the context of PBH mergers formed during early times~\cite{Sasaki:2016jop}. 

In the rest frame, for an individual halo, the probability that its nearest neighbor lies within the Yukawa effective range $l_\mathrm{y}$ is given by
\begin{equation}
    P(a<l_\mathrm{y})=1-e^{-\rho_\mathrm{h}\frac{4}{3}\pi l_\mathrm{y}^3}.
\end{equation}

Similarly, the probability that both the nearest and the second-nearest neighbors are located within the Yukawa range is
\begin{equation}
    P(a<b<l_\mathrm{y})=1-\left(1+\rho_\mathrm{h}\frac{4}{3}\pi l_\mathrm{y}^3\right)e^{-\rho_\mathrm{h}\frac{4}{3}\pi l_\mathrm{y}^3},
    \label{prob}
\end{equation}
where $\rho_{h}$ is the number density of the halos.

For those halos that are separated from each other with a distance $>l_\mathrm{y}$, the Yukawa force is screened. Consequently, these halos are coupled to the Hubble flow during radiation domination era, so that there is little chance for them to form binary during the early stage. However, they may later meet due to gravity at late times, potentially corresponding to case III in Appendix~\ref{twobody}.

Naively, we assume a fraction of those pairs we counted by \equaref{prob} can successfully form stable binary and merge during the early time in a small redshift interval $\delta z$, the event happening per unit comoving volume in the interval can be estimated by
\begin{equation}
        N(z)\delta z\approx\frac{1}{2}\rho_{\mathrm{h},z}\left[1-\left(1+\rho_\mathrm{h}\frac{4}{3}\pi l_\mathrm{y}^3\right)e^{-\rho_\mathrm{h}\frac{4}{3}\pi l_\mathrm{y}^3}\right].
\end{equation}
where 
\begin{equation}
    \rho_{\mathrm{h},z}=\frac{\rho_\mathrm{h}}{(1+z)^3}=f_{\mathrm{h}}\frac{3H_0^2}{8\pi G}\frac{\Omega_{\mathrm{DM,0}}}{m_{\mathrm{h}}}
\end{equation}
is the comoving number density of halos, and $m_h$ is the mass of the halo, $f_{\mathrm{h}}$ is the fraction of the merger remnants in DM. For the case that $f_h=10^{-10}$, $m_{\mathrm{h}}=10^{-5}M_{\odot}$ and $\rho_\mathrm{h}=1/(4/3\pi (3l_{\mathrm{y}})^3)$, it gives rise to an overall comoving number density of the events $N(z)\delta z\approx 12.3/\mathrm{Gpc}^3$.


\section{The criterion for the Halo Formation}
\label{haloform}
In the presence of a long-range attractive force much stronger than gravity, the density fluctuations of dark sector fermions can grow significantly even in the radiation-dominated era. 
Following the method of \cite{Domenech:2023afs}, we make a Newtonian fluid analysis to demonstrate the exponential growth behavior of the density perturbations due to this `fifth force'. Here we use $\beta$ to characterize the strength of the force as defined in the main text. Assuming that the annihilation and creation processes of $\psi-\bar\psi$ are frozen out in the epoch of consideration, we can treat the number density of the fermions as a conserved quantity,
\begin{equation}
    \dot n_\psi+3Hn_\psi=0,
\end{equation}
here $n_\psi$ is the number density of the fermions and $H$ is the Hubble parameter, satisfying
\begin{align}
3M_{\mathrm{pl}}^2H^2=\rho_{r}+\rho_{m}+\rho_{\psi},
\end{align}
where $\rho_{m}\propto a^{-3}$ is the matter energy density, and $\rho_{r}\propto a^{-4}$ is the radiation energy density, both excluding the contribution from DM fermions. The $
\psi$ energy density $\rho_\psi$ could evolve as either radiation or matter depending on the temperature of the universe. Here, we focus on the epoch that $\rho_\psi$ stays in the matter regime with a constant effective mass. 
The evolution of fermion density contrast is governed by a continuity equation
\begin{equation}
    \dot\delta_\psi+\theta_\psi=0,
    \label{contieq}
\end{equation}
where $\delta_\psi=\delta n_\psi/n_\psi$, and $\theta_\psi=\vec{\nabla}\vec{n_{p}}$ is the divergence of the peculiar velocity, which follows the equation
\begin{equation}
\dot\theta_{\psi}+2H\theta_\psi=-\frac{1}{a^2} \nabla^2\phi,
\label{veloeq}
\end{equation}
where $\phi$ represents the potential of the forces, including gravity and the scalar force. 
For gravitational potential, assuming the energy density of $\psi$ is the dominating source, we have
\begin{equation}
    \frac{1}{a^2}\nabla^2\phi_{\mathrm{g}}=\frac{1}{2}M_{\mathrm{pl}}\rho_\psi\delta_{\psi}
\end{equation}
On the other hand, the Yukawa potential is subject to exponential screening beyond the Yukawa length scale $l_{\mathrm{y}}$,
\begin{align}
    [\nabla^2-1/l^2_{\mathrm{y}}]\phi_{\mathrm{y}}=2\beta^2\nabla^2\phi_{\mathrm{g}}.
\end{align}

In Fourier space, the equation of motion for the number density contrast of the $\psi$ is obtained by combining \equaref{veloeq} and \equaref{contieq},
\begin{align}
& \ddot{\delta_{k}}+2H\dot{\delta_{k}}-\frac{3}{2}H^2\Omega_{\psi}\left(1+\frac{2\beta^2}{1+(kl_{\mathrm{y}})^{-2}}\right)\delta _k=0.
\end{align}
where $\Omega_{\psi}=\rho_\psi/(\rho_r+\rho_m+\rho_\psi)$ is the fraction of $\psi$, and $\Omega_{\psi}+\Omega_m+\Omega_{\mathrm{rad}}=1$, where $f_\psi=\rho_\psi/(\rho_m+\rho_\psi)$ is the fraction of $\psi$ in matter component. Deep inside of the radiation domination $H\propto a^{-2}$, the number density of the charge evolves following
\begin{equation}
    \delta_k\propto c_1 I_{0}(\sqrt{6\alpha x})+c_2  K_{0}(\sqrt{6\alpha x}),
\end{equation}
where $x=a/a_\mathrm{eq}$, $I_0$ and $K_0$ are the 0th order modified Bessel functions of the first kind and the second kind, and 
\begin{align}
    \alpha=f_\psi\left[1+\frac{2\beta^2}{1+(kl_\mathrm{y})^{-2}}\right].
\end{align}
In the limit that the force is either screened  $kl_{\mathrm{y}}\ll1$ or small $2\beta^2\ll 1$, $ I_0(\sqrt{6\alpha x})\sim 1$, so the number density is barely growing, as expected in the radiation-dominated era.
However, when the screening length scale is large, $kl_{\mathrm{y}}\gg1$, we obtain $\alpha=2f_\psi(\beta^2+1)$.
As long as $6\alpha x\gg1$, we are in the exponential growth regime since
\begin{equation}
     I_0(\sqrt{6\alpha x})\rightarrow \frac{e^{\sqrt{6\alpha x}}}{\sqrt{2\pi\sqrt{6\alpha x}}}.
\end{equation}
When $\delta_\psi$ exceeds unity, the growth of the number density of the fermions enters the non-linear regime and virialized halos begin to form~\cite{Domenech:2023afs, Savastano:2019zpr}. Notably, this exponential growth may happen much earlier than the usual gravitational structure formation, and it translates to the bound
\begin{equation}
    \frac{a_\mathrm{eq}}{a_\mathrm{i}}\approx\frac{z_\mathrm{i}}{z_\mathrm{eq}}\ll12\beta^2f_\psi.
\end{equation}
Immediately after $z_i$, the density contrast reaches the non-linear regime, and the halo dynamics after virialization are driven by radiative cooling from the emission of $\phi$ radiations.

The halos may keep accreting until the Yukawa range re-enters the horizon at $H^{-1}=l_{\mathrm{y}}$.
This constrains the maximal mass of the final halo, which is given by the total mass of $\psi$ enclosed within $l_{\mathrm{y}}$,
\begin{equation}
    M_{\mathrm{y}}= 2\pi f_{\psi}\left(\frac{a_{\mathrm{end}}}{a_{\mathrm{eq}}}\right)M_{\rm pl}^2l_{\mathrm{y}}.
\end{equation}

Assuming the final halo formation is also in the radiation-dominated era, the redshift at the end of the Yukawa structure formation is
\begin{align}
z_{\mathrm{end}}\approx\left(\Omega_{\mathrm{rad},0}H_0^2l_{\mathrm{y}}^2\right)^{-1/4}.
\end{align}
To summarize, for efficient halo formation in the radiation-dominated era, the requirements
\begin{align}   12\beta^2f_\psi z_{\mathrm{eq}}\gg {z_\mathrm{i}}\gtrsim\left(\Omega_{\mathrm{rad},0}H_0^2l_{\mathrm{y}}^2\right)^{-1/4}
\label{criterion}
\end{align}
should be satisfied, which excludes the top-right corner of the parameter space on the ($m_{\psi}$,$m_\phi$) plane in Fig. \ref{parameterspace1}.




\end{document}